\documentclass[11pt]{article}
\usepackage[]{amscd,amsfonts,amsthm}
\def\directunion{\hbox{$\bigcirc$ \hskip - 11.3 pt \raise 0.1pt
\hbox{$\scriptstyle \vee$}}\ }

\newtheorem{theorem}{Theorem}

\newtheorem{lemma}{Lemma}
\newtheorem{proposition}{Proposition}

\usepackage[psamsfonts]{amssymb}

\def\beq{\begin{equation}}
\def\eeq{\end{equation}}

\def\sep{\hbox{$\bigcirc$ \hskip - 11 pt \raise 0.1pt
\hbox{$\scriptstyle \wedge$}}\ }
\def\dperp{{\perp\perp}}

\begin{document}

\title{Linearity and compound physical systems: the case of
two separated spin 1/2 entities\footnote{Published as: D. Aerts and F. Valckenborgh, ``Linearity and compound physical systems: the case of
two separated spin 1/2 entities", in {\it Probing the Structure
of Quantum Mechanics: Nonlinearity, Nonlocality, Computation and Axiomatics}, eds. D. Aerts, M. Czachor
and T. Durt, World Scientific, Singapore (2002).}}

\author{Diederik Aerts and Frank Valckenborgh}
\date{}
\maketitle

\centerline{Center Leo Apostel (CLEA) and}
\centerline{Foundations of the Exact Sciences
(FUND),}
\centerline{Brussels Free
University, Krijgskundestraat 33,}
\centerline{1160 Brussels,
Belgium.}
\centerline{diraerts@vub.ac.be, fvalcken@vub.ac.be}


\begin{abstract}
\noindent
We illustrate some problems that are related to the
existence of an underlying linear structure at the level of
the property lattice associated with a physical system, for
the particular case of two explicitly separated spin 1/2
objects that are considered, and mathematically described, as
one compound system. It is shown that the
separated product of the property lattices corresponding with
the two spin 1/2 objects does not have an underlying linear
structure, although the property lattices associated with the
subobjects in isolation manifestly do. This is related at a
fundamental level to the fact that separated products
do not behave well with respect to the covering law (and 
orthomodularity) of elementary
lattice theory. In addition, we discuss the
orthogonality relation associated with the separated product in
general and consider the related problem of the behavior of the
corresponding Sasaki projections as partial state space mappings.
\end{abstract}

\section{Introduction}

In another contribution in this volume \cite{AV2002a}, we have
given an overview of a general mathematical framework, known under 
several names,
that can be used for the description of physical systems in
general and compound physical systems in particular. This
framework was developed in its most important aspects in Geneva and Brussels
\cite{Piron1964,Jauch1968,JP1969,Piron1976,Aerts1981,Aerts1982,Piron1990,CN1991,Aerts1994,Moore1999}.
One of the characteristics of this approach is the fact that the 
basic, primitive
elements of the formalism have a sound realistic and operational 
interpretation.
Indeed, a physical entity is described by means of its states, and 
the experimental
projects which can be performed on samples of this system. Additional 
structure is
gradually introduced as a series of physical postulates or mathematical axioms,
ranging from the physically very plausible to axioms of an admittedly more
technical nature, the latter introduced with the aim of bringing the structure
closer to standard classical and quantum physics. We want to emphasize the
generality of such an axiomatic approach and the fact that the 
results are valid in
general, independently of the particularities of the formalism.

It has been shown that two of the more technical of these axioms --- that are
definitely satisfied for standard quantum systems --- are not valid in the
mathematical model that results from these general prescriptions for a compound
physical system that consists of two operationally separated quantum objects
\cite{Aerts1981,Aerts1982,Aerts1994}. One of the two failing axioms 
is equivalent
with the linearity of the set of states for a quantum entity, hence with the
superposition principle.

One of the themes of this book is to investigate how the failure of this
``linearity" axiom is related to other perspectives on the
problem of a ``non-linear" quantum mechanics. In this paper we want 
to apply our
axiomatic approach to the particular case of two separated spin 1/2 
objects that
are described as a whole. According to standard quantum physics, an 
isolated spin
1/2 system can be mathematically represented by the complex Hilbert space
${\mathbb C}^2$. More precisely, its set of possible states
corresponds with the collection of all one-dimensional subspaces
(rays) in this space, and observables with (some of the)
self-adjoint operators on ${\mathbb C}^2$. The advantage is that for this
relatively simple situation we can not only explicitly construct a mathematical
model, but also keep an eye on the physical meaning of the mathematical
objects and understand why the linearity axiom of standard quantum mechanics
fails, at least in this case.

Let us give a brief overview of the basic ideas of the approach. In the next
section, these ideas will become more clear, when we apply them to a 
particular example,
the spin part of a single spin 1/2 object, {\it in extenso}. According to the
prescriptions of the axiomatic approach, one should first construct 
the property
lattice $\cal L$ and set of (pure) states
$\Sigma$ associated with the physical system under investigation, reflecting an
underlying program of realism that is pursued \cite{JP1969}. In 
general, the state
space is an orthogonality space,\footnote{An orthogonality space 
consists of a set
$\Sigma$ and an orthogonality relation $\perp$, that is, a relation that is
anti-reflexive and symmetric. One writes
$A^\perp = \{q \in \Sigma\ |\ q \perp p\ \hbox{{for all}}\
p \in A\}$, for
$A \subseteq \Sigma$.} while the property lattice, which is
constructed from a class of {\sl yes/no-experiments}, is always a complete atomistic lattice, usually
taken to be orthocomplemented as well
\cite{Piron1990}. The connection between both structures is
given by the Cartan map
\beq
\kappa :
{\cal L} \to {\cal P}(\Sigma) : a \mapsto \big\{ p \in
\Sigma\ |\ p \triangleleft a \big\}
\eeq
where $\triangleleft$ implements the physical idea of
actuality of $a$, if the physical system is in a state $p$.
The Cartan map is always a meet-preserving unital injection, hence
${\cal L}
\cong \kappa[{\cal L}] \subseteq {\cal P}(\Sigma)$, leading to a state space
representation of the property lattice. In addition, denoting the 
collection of all
atoms in
$\cal L$ by
$\Sigma_{\cal L}$, we have $\kappa[\Sigma_{\cal L}] = \big\{ \{p\}\ |\ p \in
\Sigma\big\} \cong
\Sigma$, hence we can identify these two sets, which we will often
do. From a physical perspective, this relation reflects the fact
that a physical state should embody a maximal amount of information
at the level of the property lattice
$\cal L$, even for individual samples of the physical system. In the axiomatic
approach, a prominent role is played by the collection of biorthogonally closed subsets
${\cal F}(\Sigma) = \big\{ A \subseteq \Sigma\ |\ A = A^\dperp \big\}$ of
$\Sigma$.  Indeed, the orthocomplementation can be introduced under the form
of two axioms, which imply that $\kappa[{\cal L}] \subseteq {\cal
F}(\Sigma)$ and $\kappa[{\cal L}] \supseteq {\cal
F}(\Sigma)$, respectively.  This {\sl state-property duality} lies at the
heart of the axiomatic approach \cite{Aerts1994,Moore1999}.

Using this general framework, one of the basic aims is to
establish a set of additional specific axioms, free from any probabilistic
notions at its most basic level, to recover the formalism of standard quantum
physics. Therefore, this approach is a theory of individual physical
systems, rather than statistical ensembles. In doing so,
a general theory is developed not only for quantal systems, but that also
incorporates classical physical systems. The classical parts of a physical
system are mathematically reflected in a decomposition of the property lattice
in irreducible components \cite{Piron1976,Aerts1981,Aerts1982,Valck1997}. For a
genuine quantum system then, that satisfies all the requirements put
forward in
\cite{Piron1976} and \cite{Aerts1981,Aerts1982}, the celebrated
representation
theorem of Piron states that these property lattices can be represented in a
suitable generalized Hilbert (or orthomodular) space. More
precisely, he showed that every irreducible complete atomistic
orthocomplemented lattice
$\cal L$ of length $\ge 4$ that is orthomodular and satisfies
the covering law (sometimes called a Piron lattice), can be
represented as the collection of all closed subspaces
${\cal L}({\cal H})$ of an appropriate orthomodular space
${\cal H}$ \cite{Piron1964}.  Mathematically speaking,
there then exists a c-isomorphism ${\cal L} 
\cong {\cal L}({\cal H})$.\footnote{A unital c-morphism
between two complete ortholattices is a mapping that
preserves arbitrary joins and orthocomplements.} The
physical motivation for this particular lattice structure
comes mainly from realistic and operational considerations. At
first sight, the mathematical demands of orthomodularity and
covering law look rather technical. They are usually
justified by taking a more active (and ideal) point of
view with respect to the physical meaning of the elements in
the property lattice (for an overview, see \cite{BC1981}).


\section{A Single Spin 1/2 System}

To illustrate the physical meaning of these mathematical
considerations, we shall treat some relatively simple particular
cases {\it in extenso}. First, we illustrate the construction of the
property lattice and state space for the
spin part of a single spin 1/2 physical system. Denote the
collection of possible states or,
alternatively, preparations, for such a physical system by
$\Sigma$.  As we have seen, empirical access to the physical
system is formalized by a set
of yes/no-experiments $\cal Q$, and we proceed with an investigation 
of $\cal Q$,
which will correspond with Stern-Gerlach experiments.

More precisely, for each spatial direction, a non-trivial definite experimental
project is associated with a Stern-Gerlach experiment in that
direction, relative to some reference direction;
$\alpha_{\theta,\phi}$ denotes the experimental project associated
with such an experiment in the direction given by
$(\theta,\phi)$, with
the following prescription for the attribution of results, if
the experiment is properly conducted on a particular sample of the 
physical system:
\begin{quote}
Attribute the positive result (outcome ``yes") if the spin
1/2 object is
detected at the upper position; otherwise,
attribute a negative result (outcome ``no").
\end{quote}
\noindent The collection of all yes/no-experiments will be denoted by $Q$. Consequently, at this point
\beq
Q \supseteq \big\{ \alpha_{\theta,\phi}\ |\ 0 \le \theta
< \pi,\ 0 \le \phi < 2\pi \big\}
\eeq
\noindent The states of the spin 1/2 particle are the spin states 
$p(\theta,\phi)$ in
the different spatial directions:
\beq
\Sigma = \big\{ p(\theta, \phi)\ \vert\ \ 0 \le \theta
< \pi,\ 0 \le \phi < 2\pi \big\}
\eeq
One of the fundamental ingredients of any physical theory is linked 
with the following
somewhat imprecise statement:
\begin{quote}
The yes/no-experiment $\alpha$ gives with certainty the outcome ``yes"
whenever the sample object happens to be in a state $p$.
\end{quote}
This statement will be expressed symbolically by a binary relation 
between the set of
states and the class of yes/no-experiments. More precisely, the 
connection between the
experimental access to the physical system and physical reality itself can be
formalized by a binary relation
$\triangleleft
\subseteq \Sigma \times Q$.
This relation symbolizes the following idea: $p \triangleleft \alpha$
means that if the physical
system is (prepared) in a state $p$, the positive result for $\alpha$
would be obtained, should one
execute the yes/no-experiment. In this case, the yes/no-experiment is said to
be {\sl true} for the object, if it is in the state
$p$. It is conceptually important to note the counterfactual 
locution. Indeed, this
formulation will allow us to attribute many properties to a 
particular sample of a
physical system. The binary relation induces in  a natural way a map, which is
intimately related to the Cartan map:
\beq
S_T : Q \to {\cal P}(\Sigma) : \alpha \mapsto \big\{ p \in
\Sigma\ |\ p
\triangleleft \alpha \big\}
\eeq
For the spin 1/2 particle it is an experimental fact that $p(\theta, \phi)
\triangleleft \alpha_{\theta',\phi'}$
iff $(\theta, \phi) = (\theta', \phi')$. There is no relation
$\triangleleft$ between $p(\theta, \phi)$ and
$\alpha_{\theta',\phi'}$ when $(\theta, \phi) \not= (\theta', \phi')$.

$Q$ is naturally equipped with an inversion relation
\beq
\widetilde{}\ : Q \to Q : \alpha \mapsto \widetilde\alpha
\eeq
the yes/no-experiment $\widetilde\alpha$ has by definition the
same experimental set-up as $\alpha$, but the
positive and negative alternatives are interchanged. This means that $p
\triangleleft \widetilde\alpha$ if the
yes/no-experiment $\alpha$ gives with certainty the outcome ``no" whenever the
state of the physical entity is $p$.
One then has the induction of a natural, physically motivated
pre-order structure on
$Q$:
\beq
\alpha < \beta\ \hbox{{iff}}\ S_T(\alpha) \subseteq S_T(\beta)
\eeq
which is used to generate the property lattice. Indeed, it is
natural to call two yes/no-experiments
equivalent if they cannot be distinguished experimentally, that is,
$\alpha \approx \beta\
\hbox{{iff}}\ S_T(\alpha) = S_T(\beta)\ \hbox{{iff}}\ p \triangleleft \alpha
\Leftrightarrow p \triangleleft \beta$. For a quantum spin 1/2 
particle, it is well
known that, according to experiment, one has
\beq
\widetilde{\alpha}_{\theta,\phi} \approx \alpha_{\pi -
\theta, \phi + \pi}
\eeq
At this moment, we have made $Q$ into a pre-ordered class, with 
some sort of an
inversion relation. There is a fundamental operation which associates with any
collection of yes/no-questions a new yes/no-experiment. Thus, the set of
yes/no-experiments should be closed under products. More formally, we 
have an operation
\beq
\Pi : {\cal P}(Q) \to Q : \big\{\alpha_j\ |\ j \in J
\big\} \mapsto
\Pi \big\{\alpha_j\ |\ j \in J \big\}
\eeq
The experimental procedure for this yes/no-experiment consists in
choosing randomly one of the $\alpha_i$ and executing the associated
experiment. With this specification, we obviously have
\beq
\widetilde{\Pi}
\big\{\alpha_j\ |\ j \in J \big\} = \Pi
\big\{\widetilde{\alpha}_j\ |\ j \in J
\big\}
\eeq
which appears somewhat strange at first, and its misunderstanding has 
been a point of
some dispute in the past. In fact, this clever definition of product 
experiments allows
us to attribute  unambiguously various different properties to (some 
preparation of) a
particular physical system, without having to explicitly test for all 
properties on the
same object
\cite{Aerts1981}. According to our prescriptions the binary relation
$\triangleleft$ should satisfy $p \triangleleft \Pi \big\{\alpha_j\ 
|\ j \in J \big\}
\Leftrightarrow p \triangleleft \alpha_j$ for all $j \in J$, or equivalently
\beq
S_T(\Pi \big\{\alpha_j\ |\ j \in J \big\}) = \bigcap
\big\{S_T(\alpha_j)\ |\ j \in J\big\}
\eeq
For example, for a spin 1/2 particle it is experimentally known that
\beq
\Pi(\{\alpha_{\theta,\phi},\alpha_{\theta',\phi'}\}) \approx \widetilde\tau
\eeq
unless $(\theta,\phi)$ and $(\theta',\phi')$ represent the same 
spatial directions.

Finally, there exist {\sl trivial} yes/no-experiments
$\tau$ and
$\widetilde\tau$. A possible experimental procedure for $\tau$ would
consist in doing nothing with the
physical system under consideration and always give the positive
result. Both yes/no-experiments are in some sense ideal elements, and 
can be viewed as
being added for technical reasons.

The equivalence relation $\approx$ on $Q$ partitions $Q$ in 
the collection of
equivalence classes, according to a standard argument. Moreover, the 
pre-ordered
structure on
$Q$ collapses into a
partial order on ${\cal L} := Q / \approx = \{ [\alpha]\ |\ 
\alpha \in Q\}$, with $[\alpha]$ denoting the equivalence class of $\alpha$;
$S_T$ lifts to the Cartan map
$\kappa$, and
$\cal L$ can be mentally put in a one-to-one correspondence with the
set of properties
or elements of reality of the physical system, in a sense derived
from that of Einstein, Podolsky and Rosen \cite{EPR1935}. Moreover,
it is not very difficult to show that
$\cal L$ becomes a complete lattice \cite{Piron1976}, with
\beq
\bigwedge \{[\alpha_j]\ |\ j \in J\} = [\Pi\{ \alpha_j\ |\ j
\in J\}]
\eeq
Given the physical meaning of the equivalence relation, one can 
unambiguously state that
a property is {\sl actual} if one of its corresponding yes/no-experiments is
true. It is the lattice $\cal L$ that we use to describe the properties of a
physical system. Note that $\cal L$, like any complete lattice, 
always contains a
maximal element $I = [\tau]$ and a minimal element $0 = [\widetilde\tau]$.

From now on, we will denote equivalence classes $[\alpha]$ by $a$. 
For our particular
example, we put $a(\theta,\phi) := [\alpha_{\theta,\phi}]$, to make 
the distinction
very clear.

Also the binary relation $\triangleleft$ lifts to the level of the 
property lattice
$\cal L$. With some abuse of notation, we then have $p \in \kappa(a) 
\Leftrightarrow p
\triangleleft a$. The physical interpretation is the following: $p 
\in \kappa(a)$
stands for ``The property $a$ is actual if the physical system is in 
a state $p$". For
our example, we have $\kappa(a(\theta, \phi)) =
\{p(\theta, \phi)\}$.

In a complete lattice, any subcollection of elements has a join or 
supremum. For a set
of properties
$([\alpha_i])_i
\in {\cal L}$ the join can be defined in a
purely mathematical way as follows:
\beq
\bigvee \{[\alpha_j]\ |\ j \in J\} = \bigwedge \{ [\beta]\ |\ 
\alpha_j < \beta\
\ \hbox{{for all}}\ j \in J \}
\eeq
It is for this reason that the join of a collection of properties has 
no obvious
physical interpretation.\footnote{We remark that the
meet operation is equivalent to logical conjunction. The ``join"
operation is however in general not equivalent to
the ``or" operation of logic. For classical physical entities the ``join"
operation is equivalent to logical disjunction, but this is not the 
case for quantum
entities. This fact is at the origin of the
common use of the word ``quantum logic" for the lattice structure that
arises in this way.}

Let us reconsider our example. Identifying, with some abuse of terminology,
the properties with their corresponding equivalence classes $a(\theta,\phi) =
[\alpha_{\theta,\phi}]$, it is true
that
\beq
a(\theta,\phi) \wedge a(\theta',\phi') = 0
\eeq
if $(\theta,\phi)$ and $(\theta',\phi')$ represent
different spatial directions, since
$\Pi\{\alpha_{\theta,\phi},\alpha_{\theta',\phi'}\} \approx
\widetilde\tau$. Indeed, there are no preparations for a spin 1/2
system in which both properties can be actual at the same time.

For our example, it is an experimental fact that if we consider two 
yes/no-experiments
$\alpha_{\theta, \phi}$ and
$\alpha_{\theta', \phi'}$ where $(\theta, \phi)$ and $(\theta', \phi')$
represent different spatial directions,
there is no third yes/no-experiment of the type $\alpha_{\theta'', \phi''}$
such that $\alpha_{\theta, \phi} <
\alpha_{\theta'', \phi''}$ and $\alpha_{\theta', \phi'} < \alpha_{\theta'',
\phi''}$. This proves that the only
yes/no-experiment $\beta$ such that $\alpha_{\theta, \phi} < \beta$ and
$\alpha_{\theta', \phi'} < \beta$ is
$\tau$. Hence for $(\theta,
\phi) \not= (\theta', \phi')$ we have:
\beq
a(\theta, \phi) \vee a(\theta', \phi') = I
\eeq

At this moment, we have found from operational considerations all the 
structural
ingredients to define the basic mathematical structure attributed to 
the compound
system that consists of two (operationally) separated spin 1/2 particles. This
structure consists in a triple
$(\Sigma,{\cal L},\kappa)$ or $(\Sigma,{\cal L},\triangleleft)$, that 
we have called a
state-property system elsewhere \cite{Aerts1999,ACVV1999}. The 
elements of $\Sigma$ are
the states attributed to the physical system under investigation, the 
elements of $\cal
L$ correspond with its possible properties, and the connection 
between both sets is
given by a Cartan map or, equivalently, a suitable binary relation, 
as we have seen.
\beq
\Sigma = \big\{ p(\theta,\phi)\ |\ 0 \le \theta
< \pi,\ 0 \le \phi < 2\pi \big\}
\eeq
\beq
{\cal L} = \{a(\theta, \phi)\ \vert\ 0 \le \theta < \pi,\ 0 \le \phi < 2\pi
\} \cup \{0\} \cup \{I\}
\eeq
\beq
\kappa(0) = \emptyset, \quad \kappa(a(\theta, \phi)) = \{p(\theta,
\phi)\},\ {\rm and}\ \kappa(I) = \Sigma
\eeq
In particular, note that $\kappa$ maps atoms in $\cal L$ to 
singletons in $\Sigma$, and
that $\kappa$ gives a state space representation of $\cal L$.

The (infinite) property lattice $\cal L$ for a
single spin 1/2
object can be visually displayed by giving its Hasse diagram:

{\hbox{\vrule height 3.4 truecm width 0pt
\special{illustration HasseSpin1/2.eps scaled 800}\vrule
height 0pt width 5truecm}}

\noindent
In the axiomatic approach, two states are defined to be {\sl
orthogonal} if there exists a yes/no-experiment $\alpha \in Q$ such
that $p \triangleleft \alpha$ and $q \triangleleft \widetilde\alpha$.
For our example, only one state will be orthogonal to a given state 
$p(\theta,\phi)$,
being the state $p(\pi - \theta, \phi + \pi)$. In this way, $\Sigma$ becomes an
orthogonality space.

It is an experimental fact that $a(\theta,\phi)$ is never
a {\sl classical} property.\footnote{A ``property" $a = [\alpha]$ is
said to be classical, if for any state of the physical system
either
$\alpha$ or $\widetilde\alpha$ is true.} Indeed, if we
prepare a spin 1/2 object in a state that corresponds to a
direction orthogonal to
$(\theta,\phi)$, then neither
$\alpha_{\theta,\phi}$ nor $\widetilde{\alpha}_{\theta,\phi}$ is
true.

For the sake of illustration, let us also consider a second
measurement scheme that would be deemed equivalent with
$\alpha_{\theta,\phi}$ according to the axiomatic approach. Let
$\beta_{\theta,\phi}$ have the same experimental arrangement, except
for the fact that the bottom channel is blocked by a suitable
absorbing device, in order to prevent an object to be localized
below. It is experimentally known that
$S_T(\alpha_{\theta,\phi}) = S_T(\beta_{\theta,\phi})$, hence
$\alpha_{\theta,\phi} \approx \beta_{\theta,\phi}$. Observe also that
$\Pi\big\{\alpha_{\theta,\phi}, I \big\} \approx \alpha_{\theta,\phi}$,
but
$\widetilde{\Pi}\big\{\alpha_{\theta,\phi}, I \big\} \not\approx
\widetilde{\alpha}_{\theta,\phi}$.

The connection between the axiomatic approach and the standard quantum
mechanical description of the spin part of a single spin 1/2 
particle, is given by the
well known (and easy to see) fact that this lattice can be 
represented as the collection
of all closed subspaces of
${\mathbb C}^2$, that is, ${\cal L}({\mathbb C}^2)$, with
\beq
a(\theta,\phi) \mapsto [(\exp(-i
\frac{\phi}{2})\cos\frac{\theta}{2},
\exp(i\frac{\phi}{2})\sin\frac{\theta}{2})]
\eeq
Note that this mapping indeed preserves the orthogonality relation.

For a single spin 1/2 object, we thus have a relatively simple property
lattice,  in which all non-trivial elements are also
representatives of (pure) states. Denoting the collection of
one-dimensional subspaces of the Hilbert space ${\cal H} = {\mathbb
C}^2$ by $\Sigma_{\cal H}$, we can also put $\Sigma \cong
\Sigma_{\cal H} \cong {\mathbb C}P^1$, this last set being
complex projective 1-space.

Once one has arrived at the basic structure of a state-property system, the
axiomatic approach proceeds by introducing further
axioms on this structure, with the aim of bringing the structure 
closer to standard
quantum mechanics. It is an easy task to verify that all the axioms, 
as stated in
\cite{AV2002a}, are satisfied for the property lattice displayed 
above. In the next
section, however, we will give an explicit example in which the axioms of
orthomodularity and the covering law both fail.


\section{The Separated Product of Two Spin 1/2 Systems}

One of the easiest compound physical systems that intuitively
and conceptually presents itself, is the
case of two separated spin 1/2 objects that are described as one
whole. Consider two such systems, respectively represented by
property lattices $ {\cal L}_i({\mathbb C}^2)$, for $i = 1, 2$,
and suppose that we want to give a mathematical description
for this situation. In this section, we will explicitly
construct the property lattice and state space that
corresponds to this physical situation.

In general, the separated product --- the mathematical
description of this situation ---
${\cal L}_1
\sep {\cal L}_2$ of
${\cal L}_1$ and
${\cal L}_2$ can be constructed in two different ways.
First, one can give an explicit construction from the
bottom up, starting from the collection of yes/no-experiments for this system.
This construction has the advantage that every property
corresponds to an equivalence class of experimental
projects, so in principle one has at one's disposal an
experimental procedure that tests for any property. Second, the
separated product can be mathematically generated through a
biorthocomplementation procedure, starting from the orthogonality space
$(\Sigma_1 \times \Sigma_2,\perp)$, with the orthogonality
relation given by
\beq
(p_1, p_2) \perp (q_1, q_2)\ \hbox{{iff}}\ \big(p_1 \perp_1
q_1\
\hbox{{or}}\ p_2 \perp_2 q_2\big)
\eeq
This construction is more
convenient from a mathematical point of view, but has the drawback
that it is a purely formal construction, which needs an {\it a
posteriori} physical interpretation. Here, we will give an overview
of the first approach, at
least for the particular case that is the main subject of
this paper. For a more detailed exposition of the general case, we refer
to \cite{Aerts1981,Aerts1982,Aerts1994}.

First, we
should be slightly more specific about what we mean with two
objects being separated. Intuitively speaking, a necessary
operational condition should be the following: it should be
possible to devise an experimental procedure, say
$e_1 \times e_2$, with outcome set $O_{e_1} \times O_{e_2}$, on the
compound system as a whole for every pair of experiments
$(e_1,e_2)$, with
$e_1$ an experiment with outcome set $O_{e_1}$ on the first object
and
similarly for $e_2$. Moreover,
whatever experiment we decide to perform on one of the objects,
should yield a result that is independent of the state of the other
object and {\it vice versa}. That is, if the compound
system is in a state such that $(x_1,x_2)$ is a possible outcome for
the experiment $e_1 \times e_2$, then the first object is in a state
such that $x_1$ is a possible result for the experiment $e_1$, and
similarly for the second object. In addition, any
experiment corresponding to one of the subobjects, can be
executed independent of the presence or absence of the other
subobject. Moreover, if an outcome is possible for an
experiment
$e_1$ to be performed on the first object, then
this outcome can be obtained irrespective of the presence or
absence of the other object. Note that this operational idea of
separation is closely related to the notion presented by
Einstein, Podolsky and Rosen \cite{EPR1935}. Also, note that there is a
big conceptual difference between the physical notions of {\sl
separation} and {\sl interaction}, the latter notion being related to the
causal structure of physical reality.

As before, we will mainly restrict ourselves to spin
measurements on a spin 1/2 object, because in this case any
experiment on one of the subobjects has only two possible
results. On the other hand, an arbitrary experiment
of the form $\alpha_1(\theta_1,\phi_1) \times
\alpha_2(\theta_2,\phi_2)$ on the compound physical system has 4
possible outcomes: $(y,y), (y,n), (n,y)$ and $(n,n)$, where for 
notational reasons we
have slightly adapted our notation. According to the prescriptions of 
the axiomatic
approach, we have to construct the collection of yes/no-experiments 
associated with all
these product experiments. First, observe that product experiments 
which have at least
one component equivalent with a trivial experiment on  the corresponding
subobject, are equivalent with either a trivial experiment on the
compound system, or
an experiment which only involves one of the subobjects. For
example, $\alpha_1(\theta_1,\phi_1) \times \tau_2$ is true iff
$\alpha_1(\theta_1,\phi_1)$ is true, with respect to the first
subobject. To make the distinction, we will put $C\alpha_1$ to be the 
experimental
project on the compound system that consists in performing the experiment
corresponding to $\alpha_1$ on the first subobject, and a similar 
convention for
the second subobject.

Thus, let us consider a product experiment
of the form $\alpha_1(\theta_1,\phi_1) \times \alpha_2(\theta_2,\phi_2)$, which
has 4 possible outcomes. With this product experiment, one can
{\it a priori} associate $2^4$ different yes/no-experiments, 
corresponding to all
subsets of the outcome set. The two trivial subsets are equivalent with
trivial yes/no-experiments on the compound system, hence will be left
out of the rest of the discussion. Temporarily abbreviating
$\alpha_i(\theta_i,\phi_i)$ by
$\alpha_i$ and $\alpha_i(\pi - \theta_i,\phi_i + \pi)$ by
$\widetilde{\alpha}_i$ for a particular direction $(\theta_i,\phi_i)$,
we will use the following
conventional notations for yes/no-experiments associated with subsets 
of a particular
form, displayed in a well-organized table form below. At the same 
time, we indicate the
corresponding inverse yes/no-experiments:

\medskip\par
\begin{center}
\begin{tabular}{|c|c|c|} \hline
& \ & \ \\
Notation & Outcome set & Inverse yes/no-experiment\\
& \ & \ \\ \hline
& \ & \ \\
$C\alpha_1$ & (y,y), (y,n) & $C\widetilde{\alpha}_1$\\
$C\alpha_2$ & (y,y), (n,y) & $C\widetilde{\alpha}_2$\\
$\alpha_1 \triangle \alpha_2$ &
(y,y) & $\widetilde{\alpha}_1 \triangledown \widetilde{\alpha}_2$\\
$\alpha_1 \triangledown \alpha_2$
& (y,y), (y,n), (n,y) & $\widetilde{\alpha}_1 \triangle \widetilde{\alpha}_2$\\
$\alpha_1 \Theta \alpha_2$ &
(y,y),(n,n) & $\widetilde{\alpha}_1 \Theta \alpha_2 \approx\ \alpha_1
\Theta \widetilde{\alpha}_2$ \\
\ & \ & \ \\ \hline
\end{tabular}
\end{center}
\bigskip\par

\noindent Observe that the notation $C\widetilde{\alpha}_j$ is 
unambiguous, in the sense
that we have $(C\widetilde{\alpha}_j) \approx C(\widetilde{\alpha}_j)$. Considering
yes/no-experiments of this general form, we can generate all yes/no-experiments
associated with the product experiment
$\alpha_1(\theta_1,\phi_1)
\times \alpha_2(\theta_2,\phi_2)$. For example, a yes/no-experiment
that tests for the result $(n,n)$ could be constructed as
$\alpha_1(\pi - \theta_1,\pi + \phi_1) \triangle
\alpha_2(\pi - \theta_2,\pi + \phi_2)$. Indeed, this yes/no-experiment would
be true if the
compound system is in a state such that $\alpha_1(\pi - \theta_1,\pi
+ \phi_1)$ is
true with respect to the first subobject, and $\alpha_2(\pi -
\theta_2,\pi + \phi_2)$
is true with respect to the second subobject.

In this way, we can obtain all properties for the compound
system that consists of two separated spin 1/2 objects, and
we can direct our attention towards the construction of the
property lattice. Because of the demand that both subobjects
are separated, we have
\beq p \triangleleft C\alpha_1(\theta_1,\phi_1) \ \hbox{{iff}}\ p_1 
\triangleleft_1
\alpha_1(\theta_1,\phi_1)
\eeq
\beq p \triangleleft C\alpha_2(\theta_2,\phi_2) \ \hbox{{iff}}\ p_2 
\triangleleft_2
\alpha_2(\theta_2,\phi_2)
\eeq
\beq p \triangleleft \alpha_1(\theta_1,\phi_1) \triangle
\alpha_2(\theta_2,\phi_2)\ \hbox{{iff}}\ p_1 \triangleleft_1
\alpha_1(\theta_1,\phi_1),\ \hbox{{and}}\ p_2 \triangleleft_2
\alpha_2(\theta_2,\phi_2)
\eeq
\beq p \triangleleft \alpha_1(\theta_1,\phi_1) \triangledown
\alpha_2(\theta_2,\phi_2)\ \hbox{{iff}}\ p_1 \triangleleft_1
\alpha_1(\theta_1,\phi_1)\ \hbox{{or}}\ p_2 \triangleleft_2
\alpha_2(\theta_2,\phi_2)
\eeq
$$ p \triangleleft \alpha_1(\theta_1,\phi_1)
\Theta
\alpha_2(\theta_2,\phi_2)\  \hbox{{iff either}}\ p_1
\triangleleft_1
\alpha_1(\theta_1,\phi_1)\ \hbox{{and}}\ p_2 \triangleleft_2
\alpha_2(\theta_2,\phi_2),\ \hskip1cm
$$
\vskip-.75cm
\beq \hbox{{or}}\ p_1
\triangleleft_1
\alpha_1(\pi - \theta_1,\pi + \phi_1)\ \hbox{{and}}\ p_2
\triangleleft_2
\alpha_2(\pi - \theta_2,\pi + \phi_2)
\eeq
It is not very difficult to see from these prescriptions that the 
state of the global
system is completely known whenever one knows the states of the two
separated spin 1/2 objects that make up the compound system.
Consequently, the set of states can be taken as $\Sigma_1
\times \Sigma_2$, with
\beq
\Sigma_1 = \big\{ p_1(\theta_1,\phi_1)\ |\ 0 \le \theta_1
< \pi,\ 0 \le \phi_1 < 2\pi \big\}
\eeq
\beq
\Sigma_2 = \big\{ p_2(\theta_2,\phi_2)\ |\ 0 \le \theta_2
< \pi,\ 0 \le \phi_2 < 2\pi \big\}
\eeq
However, for notational reasons we shall often use abbreviations of 
the form $p_j$ for
a general element of $\Sigma_j$.

Consequently, we can represent the property lattice corresponding to
this situation as a subcollection of ${\cal P}(\Sigma_1 \times
\Sigma_2)$. Properties of the first kind would be represented by
singletons $(p_1(\theta_1,\phi_1),
p_2(\theta_2,\phi_2))$; properties of the second kind consist of
all sets of the general form $\{p_1(\theta_1,\phi_1)\} \times
\Sigma_2 \cup \Sigma_1 \times \{p_2(\theta_2,\phi_2)\}$; finally,
properties of the third kind are two-element sets of the
general form
$\{
(p_1(\theta_1,\phi_1),p_2(\theta_2,\phi_2)),(p_1(\pi
- \theta_1,\pi + \phi_1),p_2(\pi - \theta_2,\pi + \phi_2))
\}$. The full property lattice is then generated by taking arbitrary
products of
corresponding yes/no-experiments, which amounts to taking intersections of the
corresponding images of the Cartan map, as we have seen before,
because the Cartan map is
one-to-one and preserves intersections, due to the corresponding
property of the map
$S_T$. Denoting
$S^2 := [0,\pi)
\times [0,2\pi)$, we then obtain for the full property lattice
corresponding to this physical
situation the intersection system ${\cal I}(\Omega)$
generated by the collection of all these sets $\Omega$, and
it is geometrically clear, by considering the set $\Sigma_1 \times
\Sigma_2$, that the
second equation also holds:
\begin{eqnarray}
{\cal L}_1
\sep {\cal L}_2 & \cong & {\cal I}({\cal T} \cup {\cal
A}_1
\cup {\cal A}_2 \cup {\cal S} \cup {\cal U} \cup {\cal B})\\
& \cong & {\cal T} \cup {\cal
A}_1
\cup {\cal A}_2 \cup {\cal S} \cup {\cal U} \cup {\cal P}
\end{eqnarray}
with
{\small
$$ {\cal T} = \big\{ \emptyset, \Sigma_1 \times \Sigma_2 \big\}
$$
$$
{\cal A}_1 = \big\{\{p_1(\theta_1,\phi_1)\} \times \Sigma_2\ |\
(\theta_1,\phi_1) \in S^2
\big\}\\
$$
$$
{\cal A}_2 = \big\{ \Sigma_1 \times \{p_2(\theta_2,\phi_2)\}\ |\
(\theta_2,\phi_2) \in S^2
\big\}\\
$$
$$
{\cal S} = \big\{ (p_1(\theta_1,\phi_1),p_2(\theta_2,\phi_2))\ |\
(\theta_i,\phi_i) \in S^2
\big\}
$$
$$
{\cal U} = \big\{ \{p_1(\theta_1,\phi_1)\} \times
\Sigma_2 \cup \Sigma_1 \times \{p_2(\theta_2,\phi_2)\}\ |\
(\theta_i,\phi_i) \in S^2 \big\}
$$
$$
{\cal B} = \big\{ \{
(p_1(\theta_1,\phi_1),p_2(\theta_2,\phi_2)),(p_1(\pi
- \theta_1,\pi + \phi_1),p_2(\pi - \theta_2,\pi + \phi_2))
\}\ |\ (\theta_i,\phi_i) \in S^2 \big\}
$$
$$
{\cal P}  =
\big\{(p_1(\theta_1,\phi_1),p_2(\theta_2,\phi_2)),
(p_1(\theta_1',\phi_1'),p_2(\theta_2',\phi_2'))\ |\ \theta_i \not=
\theta_i', \phi_j
\not= \phi_j' \big\}$$
}

\noindent Note that ${\cal B} \subset {\cal P}$. In this way,
we have obtained a collection of properties that together
make up the property lattice that describes a compound
system consisting of two separated spin 1/2 objects, the infimum of a 
collection of
properties being their intersection.

Some of these properties appear familiar, given
the fact that the global system consists of two subsystems of which the
mathematical description is known. On the other hand, there
are also some properties, notably in $\cal P$, which have a
more classical appearance, in the sense that they consist of
a set theoretical union of two states, without any new
superpositions that do arise in this particular way. It follows from 
geometrical
considerations that these elements arise from intersections of 
elements in ${\cal A}_1$
and
${\cal A}_2$. On the other hand, taking two different elements of
$\Sigma_1
\times \Sigma_2$ such that one of the coordinates coincides, the
property $a$ generated by
these two elements has $\pi_j(a) = \Sigma_j$, with $\pi_j$ the
canonical projection on the
other coordinate, hence contains plenty of other elements of $\Sigma_1 \times
\Sigma_2$ than the
two generating states. This situation is reminiscent to the notion of a
superselection rule in standard quantum mechanics, to which we come 
back later. Observe
that all these properties, even the more enigmatic ones, have a clear 
operational
meaning, in the sense that there exists a corresponding experimental 
procedure that can
test for this property.

What about the orthogonality relation? Suppose that $(p_1,p_2),
(q_1,q_2)
\in \Sigma_1
\times \Sigma_2$. If $p_1 \perp_1 q_1$, we have seen that there is
some direction $(\theta_1,\phi_1)$ such that $p_1$ is represented by
$a_1(\theta_1,\phi_1)$ and $q_1$ by
$a_1(\pi - \theta_1,\phi_1 + \pi)$. Because the corresponding
experiments $C\alpha_1(\theta_1,\phi_1)$ and $C\alpha_1(\pi - \theta_1,\phi_1 +
\pi)$ can also be performed on the compound system, it follows that 
$(p_1,p_2) \perp
(q_1,q_2)$. Conversely, if
$(p_1,p_2) \perp (q_1,q_2)$, there exists a yes/no-experiment $\alpha
\in Q$ such that
$(p_1,p_2) \triangleleft \alpha$ and $(q_1,q_2) \triangleleft
\widetilde{\alpha}$. Recall that $(p_1,p_2)
\triangleleft \alpha$ formalizes the physical idea that the
execution of the experiment $\alpha$ should yield a positive
result with certainty, would the experiment be properly
performed on a particular sample of the compound physical system that 
happens to be in
a state given by $(p_1,p_2)$.

There are several basic forms to be considered for
$\alpha$. Suppose first that $\alpha = C\alpha_1$ for some $\alpha_1 \in
Q_1$; we have seen that $\widetilde{\alpha} = C\widetilde{\alpha}_1$, hence
$p_1 \perp_1 q_1$. The same type of argument works if
$\alpha = C\alpha_2$ for some $\alpha_2 \in
Q_2$ to conclude that $p_2 \perp_2 q_2$. Next, suppose
that
$\alpha =
\alpha_1 \triangle \alpha_2$, then $\widetilde\alpha =
\widetilde{\alpha}_1\; \triangledown \widetilde{\alpha}_2\;$. By the
prescription of the experimental procedure and by the separation
of the two objects, formally encoded in (20) - (24), we have on the one hand,
$p_1 \triangleleft_1
\alpha_1$ and $p_2 \triangleleft_2 \alpha_2$, and on the
other hand $q_1
\triangleleft_1
\widetilde{\alpha}_1$ or $q_2 \triangleleft_2
\widetilde{\alpha}_2$, which implies that $p_1 \perp_1 q_1$ or $p_2
\perp_2 q_2$. All other cases being similar, we
conclude that
\beq
(p_1,p_2) \perp
(q_1,q_2)\ \hbox{{iff}}\ p_1 \perp_1 q_1\ \hbox{{or}}\ p_2
\perp_2 q_2
\eeq
The same type of argument also shows that the property lattice is
orthocomplemented. The reader can verify that orthocomplements
for the elements of various forms in ${\cal L}_1({\mathbb C}^2) \sep {\cal
L}_2({\mathbb C}^2)$ are those that are given in the following
convenient table.
In this table, the $p_i$ and $q_j$ stand for points
in the state spaces that correspond to the two different
subobjects. In addition, we demand that $p_1 \not= q_1$ and $p_2 \not= q_2$.

\medskip\par
\begin{center}
\begin{tabular}{|c|c|} \hline
& \ \\
Element & Orthocomplement\\
& \ \\ \hline
& \ \\
$\{(p_1,p_2)\}$ & $\{p_1\}^\perp \times \Sigma_2 \cup \Sigma_1
\times \{p_2\}^\perp$
\\
$\{(p_1,p_2), (q_1,q_2)\}$ & $\{(p_1^\perp,q_2^\perp),
(q_1^\perp,q_2^\perp)\}$
\\
$\{p_1\} \times \Sigma_2$ & $\{p_1\}^\perp \times \Sigma_2$
\\
$\Sigma_1 \times \{p_2\}$ & $\Sigma_1 \times \{p_2\}^\perp$
\\
$\{p_1\} \times \Sigma_2 \cup \Sigma_1 \times \{p_2\}$ &
$\{(p_1^\perp,p_2^\perp)\}$
\\
\ & \ \\ \hline
\end{tabular}
\end{center}
\bigskip\par

\noindent
Next, we will show that both the orthomodularity and the
covering law fail, taking the mathematical representation for our 
compound physical
system to be
${\cal L}_1({\mathbb C}^2)
\sep {\cal L}_2({\mathbb C}^2)$, which can be taken, as we have seen, 
as the property
lattice for our compound system. We start with orthomodularity. Denoting by
$[x]$ the one-dimensional subspace spanned by an element $x \in 
{\mathbb C}^2$, let $a
=
\{([\psi_1], [\psi_2])\}$ and
$b = \{ ([\psi_1], [\psi_2]), ([\phi_1], [\phi_2]) \}$, with 
$[\psi_1] \not\perp_1
[\phi_1]$ and $[\psi_2] \not\perp [\phi_2]$. Then $a^\perp = \{[\psi_1]\}^\perp
\times
\Sigma_2 \cup \Sigma_1 \times \{[\psi_2]\}^\perp$. Consequently, meets
corresponding to intersections, we have $b
\wedge a^\perp = \emptyset$, hence $a = a \vee (b \wedge a^\perp) < 
b$ (a strict
inequality!). If orthomodularity were valid, we would have obtained $a \vee
(b \wedge a^\perp) = b$, which proves our assertion.

It is also easy to show that the covering law cannot be valid
for this particular example, too. To see this, take a lattice element of 
$\cal P$, say
$\{ ([\psi_1], [\psi_2]), ([\phi_1], [\phi_2]) \}$.
It is always possible to choose a third element
$([\xi_1],[\xi_2])$, such that $\xi_1$ and $\psi_1$ are
two linearly independent elements, and also $\xi_2$ and $\phi_2$. Then
\beq
\big\{ ([\psi_1],[\psi_2]), ([\phi_1], [\phi_2]) \big\} \wedge
\{([\xi_1],[\xi_2])\} = \emptyset
\eeq
\beq
\big\{ ([\psi_1],[\psi_2]), ([\phi_1], [\phi_2]) \big\} \vee
\{([\xi_1],[\xi_2])\} = \Sigma_1 \times \Sigma_2
\eeq
This element should cover $\{ ([\psi_1], [\psi_2]), ([\phi_1], [\phi_2])
\}$ if the covering law were valid. However, the element $\{[\psi_1]\} \times
\Sigma_2 \cup \Sigma_1 \times \{[\phi_2]\}$ belongs to ${\cal
L}_1({\mathbb C}^2) \sep {\cal
L}_2({\mathbb C}^2)$ and
\beq
\big\{ ([\psi_1],[\psi_2]), ([\phi_1], [\phi_2]) \big\} \subset 
\{[\psi_1]\} \times
\Sigma_2 \cup \Sigma_1 \times \{[\phi_2]\} \subset \Sigma_1 \times \Sigma_2
\eeq
which is a contradiction, because these are strict inclusions.

We can then safely conclude that the property lattice ${\cal 
L}_1({\mathbb C}^2)
\sep {\cal L}_2({\mathbb C}^2)$ is {\sl not} isomorphic to a Piron
lattice (associated with an orthomodular space), due to the
fact that orthomodularity and the covering law fail. Consequently,
an underlying linear structure such that ${\cal L}_1({\mathbb C}^2)
\sep {\cal L}_2({\mathbb C}^2)$ would correspond to the complete lattice of all
closed subspaces is out of the question: one cannot
construct an underlying Hilbert space for which the collection of
all closed subspaces would correspond with the property lattice associated with
this physical situation.


\section{The Orthogonality Relation}

In this section, we want to take a closer look at the
orthogonality relation on a general $\Sigma_1 \times \Sigma_2$ that
generates the property lattice corresponding to the
separated product. It will be convenient to demonstrate some
general results, the first for a general orthogonality space, the 
second valid for
the particular orthogonality relation given by (19).

\begin{lemma} In an arbitrary orthogonality space
$(\Sigma,\perp)$, we have
\beq
\big(\ \bigcup_{j \in J} M_j\ \big)^\perp = \bigcap_{j \in J} M_j^\perp
\eeq
\end{lemma}

\noindent Proof:
Recall that an orthogonality relation is by definition
irreflexive and symmetric. If $A \subseteq B$, then $B^\perp
\subseteq A^\perp$,
hence $\big(\ \cup_{j \in J} M_j\ \big)^\perp \subseteq M_k^\perp$,
for each $k \in J$.
Observe also that $A \subseteq A^\dperp$ for any $A \subseteq
\Sigma$, by symmetry.
Consequently, if $F$ is any subset of $\Sigma$, we obtain $F \subseteq
\cap_{j \in J} M_j^\perp$ iff $F \subseteq M_j^\perp$ for each $j \in
J$ iff $M_j
\subseteq F^\perp$ for each $j \in J$ iff $\cup_{j \in J} M_j
\subseteq F^\perp$ iff
$F \subseteq \big(\ \cup_{j \in J} M_j\ \big)^\perp$, which proves
the other inclusion.
\qed

\begin{proposition}  Suppose that $\Sigma_1 \times \Sigma_2$
is an orthogonality
space, equipped with the orthogonality relation (19). Let
$M_j
\subseteq
\Sigma_j$,
$j=1,2$ and
$(p_1,p_2)
\in
\Sigma_1
\times \Sigma_2$. Then
\begin{eqnarray}
\{(p_1,p_2)\}^\perp & = & (\{p_1\}^\perp \times \Sigma_2)\ \cup\
(\Sigma_1 \times \{p_2\}^\perp)\\
(\{p_1\} \times M_2)^\perp & = & (\{p_1\}^\perp \times \Sigma_2)\
\cup\ (\Sigma_1 \times M_2^\perp)\\
(M_1 \times M_2)^\perp & = & (M_1^\perp
\times \Sigma_2)\ \cup\ (\Sigma_1 \times
M_2^\perp)
\end{eqnarray}
\end{proposition}

\noindent Proof:
First, $(r_1,r_2) \perp (p_1,p_2)$ iff $r_1
\perp_1 p_1$ or
$r_2 \perp_2 p_2$ iff $(r_1,r_2) \in \{p_1\}^\perp \times \Sigma_2$
or $(r_1,r_2)
\in \Sigma_1 \times \{p_2\}^\perp$. Second, if $r_1 \in
\{p_1\}^\perp$ or $r_2 \in
M_2^\perp$, then $(r_1,r_2) \in (\{p_1\} \times M_2)^\perp$; conversely, let
$(r_1,r_2) \in (\{p_1\} \times M_2)^\perp$; if $r_1 \in
\{p_1\}^\perp$, there is
nothing to prove; if not, take an arbitrary $m_2 \in M_2$; because
$(r_1,r_2) \perp
(p_1,m_2)$ and $r_1 \not\perp_1 p_1$, it follows that $r_2 \perp_2
m_2$, hence $r_2
\in M_2^\perp$. The final equation follows from the next
calculation, using some of the
previous results:
\begin{eqnarray*}
(M_1 \times M_2)^\perp & = & \big(\bigcup_{r_1 \in M_1} (\{r_1\} \times
M_2)\ \big)^\perp\\
& = & \bigcap_{r_1 \in M_1} \big(\ (\{r_1\} \times M_2)^\perp\ \big)\\
& = & \bigcap_{r_1 \in M_1} \big(\ (\{r_1\}^\perp \times \Sigma_2)\
\cup\ (\Sigma_1 \times M_2^\perp)\ \big)\\
& = & \big(\ \bigcap_{r_1 \in M_1} (\{r_1\}^\perp \times \Sigma_2)\ \big)\
\cup\ (\Sigma_1 \times M_2^\perp)\\
& = & \big(\ \bigcup_{r_1 \in M_1} \{r_1\}\ \big)^\perp \times \Sigma_2\
\cup\ (\Sigma_1 \times M_2^\perp)\\
& = & (M_1^\perp
\times \Sigma_2)\ \cup\ (\Sigma_1 \times M_2^\perp)
\end{eqnarray*}
what was to be proved.
\qed

Let $(\Sigma_j,
\perp_j)$,
$j = 1, 2$, be two
$T_1$ orthogonality spaces, that is, we additionally demand that 
$\forall p_j \in
\Sigma_j :
\{p_j\}^{{\perp_j}{\perp_j}} = \{p_j\}$. Suppose that there exist
$p_1, q_1 \in \Sigma_1$ such that $p_1 \not= q_1$, and
similarly for
$\Sigma_2$. The following straightforward
calculation shows that the two-element set
$\{(p_1, p_2), (q_1, q_2)\}$ is always a closed subspace of the
orthogonality space
$(\Sigma_1 \times \Sigma_2, \perp)$, with the orthogonality given by (19):
\begin{eqnarray*}
\big\{(p_1, p_2), (q_1, q_2)\big\}^\dperp & = &
\big(\big\{(p_1, p_2)\big\}^\perp \cap \big\{(q_1,
q_2)\big\}^\perp\big)^\perp\\
& = & \big((\{p_1\}^\perp \times \Sigma_2) \cup (\Sigma_1
\times \{p_2\}^\perp) \cap\\
& & \hskip.5cm \cap\ (\{q_1\}^\perp_1 \times
\Sigma_2) \cup\
(\Sigma_1
\times \{q_2\}^\perp)\big)^\perp\\
& = & \big(((\{p_1\}^\perp \cap
\{q_1\}^\perp) \times \Sigma_2) \cup (\{p_1\}^\perp
\times \{q_2\}^\perp)
\cup\\
& & \hskip.5cm \cup\ (\Sigma_1 \times (\{p_2\}^\perp \cap
\{q_2\}^\perp)) \cup (\{q_1\}^\perp
\times \{p_2\}^\perp) \big)^\perp\\
& = & ((\{p_1\}^\perp \cap
\{q_1\}^\perp) \times \Sigma_2)^\perp \cap
(\{p_1\}^\perp
\times \{q_2\}^\perp)^\perp
\cap\\
& & \hskip.5cm \cap\ (\Sigma_1 \times (\{p_2\}^\perp \cap
\{q_2\}^\perp))^\perp \cap (\{q_1\}^\perp
\times \{p_2\}^\perp)^\perp\\
& = & (\{p_1, q_1\}^\dperp
\times
\Sigma_2)
\cap (\{p_1\} \times \Sigma_2 \cup \Sigma_1 \times \{q_2\})
\cap\\
& & \hskip.5cm \cap\ (\Sigma_1 \times
\{p_2,q_2\}^\dperp) \cap (\{q_1\} \times
\Sigma_2 \cup \Sigma_1 \times \{p_2\})\\
& = & (\{p_1\} \times \Sigma_2 \cup \{p_1,
q_1\}^\dperp \times \{q_2\}) \cap\\
& & \hskip.5cm \cap\ (\{q_1\} \times
\{p_2,q_2\}^\dperp \cup\
\Sigma_1 \times \{p_2\})\\
& = &\{(p_1, p_2)\}  \cup \{(q_1,q_2)\}\\
& = & \{(p_1, p_2), (q_1,q_2)\}
\end{eqnarray*}
Consequently, these two elements do not generate an
irreducible projective plane. So in general there exist in the
property lattice corresponding to
the separated product, a host of two-element sets that form closed 
subspaces,
relative to this
orthogonality relation, a situation that is unheard off in
standard quantum physics.

In a usage derived from that of standard quantum physics, one can say
that two properties $a$ and $b$ in a property lattice are separated by a {\sl
superselection rule} whenever $p \triangleleft a \vee b$ implies either $p
\triangleleft a$ or $p
\triangleleft b$. In standard quantum physics, all known superselection
rules can be accommodated for by restricting some global Hilbert 
space, attributed
to the physical system under investigation, to a suitable collection of
mutually orthogonal subspaces, not allowing states that do not belong to one of
these orthogonal components. However, observe that for the separated 
product of two
spin 1/2 objects there do even exist {\sl non-orthogonal} states that 
are separated
by a superselection rule, in particular pairs of states that 
constitute many of the
properties in $\cal P$.


\section{Sasaki Regularity}

Yet another characterization of the covering law can be
formulated for (complete) atomistic orthomodular lattices, using the 
projections in a
suitable involution semigroup of mappings associated with the 
property lattice. Let
$\cal L$ be a complete atomistic orthomodular lattice, with 
orthocomplementation $a
\mapsto a^\perp$, then $\cal L$ satisfies the covering law
iff each so-called Sasaki projection
\beq
\phi_a : {\cal L} \to {\cal L}: x
\mapsto (x \vee a^\perp) \wedge a
\eeq
maps any atom not smaller than $a^\perp$ to an atom, that
is, for any $a \in {\cal L}$ the restriction and
corestriction
\beq
\phi_a : \Sigma
\setminus \{ p \in \Sigma\ |\ p < a^\perp\} \to \Sigma : p
\mapsto (p \vee a^\perp) \wedge a
\eeq
is well-defined \cite{Piron1976}. In general, it is convenient to give a
special name to all Sasaki projections that satisfy this
last condition. We will call them {\sl regular} Sasaki
projections.

Because
$\cal L$ is isomorphic with the orthomodular lattice of all
Sasaki projections under some suitable
conditions \cite{BC1981}, and the Sasaki projections can be interpreted as
representing state transitions corresponding to a positive response 
for idealized
measurement procedures associated with the properties, this procedure 
refers to a
more active point of view on physical systems. Indeed, one assumes the
existence of an ideal class of measurement procedures, such that the state
before such a measurement becomes a well-defined state after the
experiment, whenever one has obtained the positive result.
In view of this interpretation, it seems indeed more natural to
consider Sasaki projections as partially defined state space mappings.
Given the fact that $\kappa[{\cal L}] = {\cal F}(\Sigma)$ under
the usual orthocomplementation axioms of the axiomatic approach,
we then have to consider a family of mappings
\beq
\phi_M : {\cal D}(\phi_M) \to \Sigma : p \mapsto
(\{p\}
\cup M^\perp)^\dperp
\cap M
\eeq
with ${\cal D}(\phi_M) \subseteq \Sigma$, and $M = \kappa(a)$
for some $a \in {\cal L}$. The latter condition arises because it is 
exactly subsets
of this form that represent properties attributed to the physical 
system. As we have
seen,
$\phi_M$ is regular iff
${\cal D}(\phi_M) = \{ p \in \Sigma\ |\ p \not\in M^\perp\}$.

Given the role of the covering law in the
representation theorems and its interpretation, it is then of 
considerable interest to
investigate the presence of any aberrant Sasaki projections for 
operationally separated
objects that are described as one compound physical system, both in 
general and with
respect to our example. Because of their putative interpretation as state
transitions, we consider the Sasaki projections as partial state 
space mappings, and
investigate them at the level of the state space description
$(\Sigma_1 \times \Sigma_2, \perp)$.

Consequently, let
$(\Sigma_1,\perp_1)$ and $(\Sigma_2,\perp_2)$ be two Sasaki regular
$T_1$ orthogonality spaces, in the sense that Sasaki projections 
associated with
biorthogonally closed sets are regular, and take
$(p_1,p_2)
\in \Sigma_1 \times \Sigma_2$ such
that $(p_1,p_2)
\not\perp M_1 \times M_2$, with $M_1 = M_1^\dperp$ and $M_2 = M_2^\dperp$.
According to our previous results, this implies that $p_1
\not\perp_1 M_1$ and $p_2 \not\perp_2 M_2$. After some calculation
efforts, one obtains
\begin{eqnarray*}
&\phi_{M_1 \times M_2}(p_1,p_2) =  \big(
\{(p_1,p_2)\} \cup (M_1 \times M_2)^\perp \big)^{\perp\perp}
\cap (M_1 \times M_2)\\
& =  \big(
\{(p_1,p_2)\} \cup (M_1^\perp \times \Sigma_2) \cup (\Sigma_1 \times
M_2^\perp)
\big)^{\perp\perp}
\cap (M_1 \times M_2)\\
& = \big( (\{p_1\}^\perp \times \Sigma_2 \cup \Sigma_1
\times \{p_2\}^\perp) \cap (M_1 \times M_2) \big)^\perp
\cap (M_1
\times M_2)\\
& = \big( (\{p_1\}^\perp \cap M_1) \times M_2 \cup M_1
\times (\{p_2\}^\perp \cap M_2) \big)^\perp
\cap (M_1 \times M_2)\\
& = \big( (\{p_1\}^\perp \cap M_1)^\perp \times \Sigma_2\
\cup\
\Sigma_1 \times M_2^\perp \big) \cap\\
&  \hskip.5cm \cap\ \big( M_1^\perp \times \Sigma_2\ \cup \
\Sigma_1 \times (\{p_2\}^\perp \cap M_2)^\perp \big) \
\cap\ (M_1 \times M_2)\\
& = \big( (\{p_1\}^\perp \cap M_1)^\perp \times \Sigma_2\
\cup\ \Sigma_1 \times M_2^\perp \big) \cap\ \\
&  \hskip.5cm \cap\ M_1 \times \big((\{p_2\}
\cup M_2^\perp)^\dperp \cap\ M_2\big)\\
& =  \big((\{p_1\} \cup M_1^\perp)^\dperp
\cap M_1\big) \times \big( (\{p_2\} \cup M_2^\perp)^\dperp
\cap M_2\big)
\end{eqnarray*}
and the right hand side belongs to $\Sigma_1 \times \Sigma_2$, by
assumption. In particular, with some abuse of notation
\beq
\phi_{(q_1,q_2)}(p_1,p_2) = (q_1,q_2)
\eeq
\beq
\phi_{\{q_1\} \times \Sigma_2}(p_1,p_2) = (q_1,p_2)
\eeq
Consequently, regularity is preserved for all Sasaki projections corresponding
to biorthogonally
closed subsets of the general form
$M_1 \times M_2$. Therefore, we have to screen for other candidates
that could violate our
regularity condition. Luckily, we don't have to look too far. Indeed,
consider one of the
peculiar biorthogonally closed sets of the form
$M = \big\{(q_1,q_2),(r_1,r_2)\big\}$, which, as we have seen, can be 
found in any
property lattice corresponding to a separated product. We will
show that, for
$(p_1,p_2) \not\in M^\perp$:
\beq
\phi_M(p_1,p_2) = M = \big\{(q_1,q_2),(r_1,r_2)\big\}
\eeq
whenever $p_1 \not\perp_1 q_1, p_1
\not\perp_1 r_1, p_2 \not\perp_2 q_2, p_2 \not\perp_2 r_2$. Indeed, 
if $(p_1,p_2)
\not\in M^\perp$, then either (1) $p_1 \not\perp_1 q_1$ and $p_2 
\not\perp_2 q_2$,
or (2) $p_1 \not\perp_1 r_1$ and $p_2 \not\perp_2 r_2$, or (3) both. 
Consequently,
\begin{eqnarray*}
&\big( \{(p_1,p_2)\} \cup M^\perp \big)^\dperp  = 
\big( \{(p_1,p_2)\}^\perp \cap\ M \big)^\perp\\
& =  \big( \{(q_1,q_2),(r_1,r_2)\} \cap\ (\{p_1\}^\perp \times
\Sigma_2\ \cup\ \Sigma_1 \times \{p_2\}^\perp )
\big)^\perp\\
& =  \left\{\matrix{ \Sigma_1 \times \Sigma_2  &  \hbox{{if (1) and (2) are
valid}}
\cr  \{q_1\}^\perp \times \Sigma_2 \cup \Sigma_1 \times \{q_2\}^\perp &
\hbox{{if only (2) is
valid}}
\cr  \{r_1\}^\perp \times \Sigma_2 \cup \Sigma_1 \times \{r_2\}^\perp &
\hbox{{if only (1) is
valid}} \cr}\right.
\end{eqnarray*}
from which (42) easily follows. Because $M$ is not an atom, $\phi_M$ is not
regular, and $\phi_M$ can no longer be interpreted as a state 
transition resulting
from a positive response for the yes/no-experiment that corresponds 
with $M$. This is
apparently due to the construction of the yes/no-experiments and the
properties associated with product experiments, and is a 
manifestation of the symmetry
with respect to the possible state transitions the two separated 
constituents can
undergo for one and the same positive outcome, attributed to the corresponding
yes/no-experiment.

In summary, if both
$\Sigma_1$ and
$\Sigma_2$ contain at least two different states and if in both state 
spaces there
exists a state that is not orthogonal to both these states, that is, for all
non-trivial orthogonality spaces, the previous argument is valid and we have
demonstrated the following

\begin{theorem} If
$(\Sigma_1,\perp_1)$ and $(\Sigma_2,\perp_2)$ are two Sasaki regular
$T_1$ orthogonality spaces, the orthogonality space $(\Sigma_1 \times
\Sigma_2,\perp)$, with the orthogonality given by (19), is not Sasaki regular,
whenever $\perp_1$ and
$\perp_2$ are non-trivial.
\end{theorem}

\noindent Because the property lattice attributed to a classical 
physical system
usually corresponds with the collection of all subsets of some set 
$\Sigma$, one
easily sees that the orthogonality relation in this case becomes trivial: every
pair of distinct states is orthogonal. Consequently, the theorem is not valid
whenever at least one of the two orthogonality spaces represents a classical
physical system.

Of course, it then also follows for the same reason
that in our particular example Sasaki regularity is not preserved.


\section{Discussion}

Several combinatorial mathematical constructions have been
proposed for the description of compound physical systems,
starting from the representations of the hypothetical
subobjects in those compound physical systems. Two of them
were studied by one of the authors: the so-called separated product, that
constructs the property lattice for the compound system that
consists of two explicitly separated physical objects
\cite{Aerts1981,Aerts1982}; the coproduct, that generates the
property lattice of two separated physical systems, for
which only experimental projects on one of the subobjects at
a time, chosen at random, are taken into account
\cite{Aerts1984,Valck2000,Valck2001}.
The name of the latter comes from the fact that one can show that it
corresponds with (the underlying object of) the mathematical coproduct in an
appropriate categorical sense.  In addition, in \cite{AD1978b}, the property lattice
associated with
the more traditional Hilbert tensor product space representation for compound
physical systems has been studied.

In this paper, we have examined in some detail the problem of the
mathematical description of the conceptually important physical situation that
consists in two separated spin 1/2 objects that are considered as one compound
physical system. In particular, we have shown that a representation as a
collection of closed subspaces of a linear space is impossible in 
general, due to
the fact that the covering law fails. Thus, there seems to be some 
relation between
the notion of a {\sl compound} physical system and the mathematical property of
{\sl linearity}. We have also spent some time on studying another 
perspective which
is intimately related to the covering law: the regularity of
the corresponding collection of Sasaki projections. It is tempting to 
speculate how
the possible development of a generalized ``non-linear" quantum physics could
eventually put in a different light the problems that quantum mechanics
experiences to describe (operationally) separated quantum objects.

For atomistic lattices, one can show that the covering law
is equivalent with the so-called {\sl exchange property},
which states that for each $x \in {\cal L}$ and for any pair of atoms
$p, q \in \Sigma_{\cal L}$, the following condition is
valid: $p \wedge x = 0$ and $p < q \vee x$ together imply
$q < p \vee x$ \cite{MM1970,FF2000}. This
condition is reminiscent of the superposition principle of
quantum mechanics (and is trivially satisfied for the
property lattice of a classical physical system, which is
usually taken to be the collection of all subsets of state
space). Consequently, the covering law seems
deeply related to the possibility of attributing a linear
structure to the state space of an arbitrary physical
system.

The fact that it is mainly the covering law that is
responsible for a linear representation of the property
lattice, also follows from the following theorem \cite{MM1970}: For
any irreducible
complete atomistic orthocomplemented lattice of length $\ge 4$ that satisfies
the covering law, there exists a division ring $K$ with an
involution
$\lambda
\mapsto \lambda^*$, and a vector space $V$ over $K$ with a
hermitian form $f : V \times V \to K$, such that $\cal L$ is
ortho-isomorphic to the lattice of all closed subspaces of
$V$, relative to
$f$. Consequently, the stronger condition of orthomodularity
is not necessary to obtain a linear
structure.\footnote{Actually, there exists an even weaker
representation theorem, which states that a linear
representation holds for irreducible complete lattices $\cal
L$ of length $\ge 4$, if both $\cal L$ and its opposite
lattice ${\cal L}^*$ are atomistic and satisfy the covering
law \cite{MM1970}.}

In our opinion, the mathematical description of
this physical situation has at least some
relevance with respect to the enigmatic classical limit. At the very
least, this approach yields another perspective on the problematic
associated with the one and the many, as it was aptly called by one of
the authors \cite{Aerts1981}. Indeed, the construction in this 
particular case seems
to be empirically and operationally adequate in that it incorporates 
all experiments
one can possibly perform on two separated spin 1/2 objects separately and as a
whole, and therefore seems to evade the critique of Cattaneo
and Nistic\'o \cite{CN1990}.

Of course, if two physical objects are separated in the operational sense that
was used in this paper, one usually does not bother about representing the
properties that explicitly account for the separation.
     From this point of view, the so-called coproduct property lattice arises
if one considers the
collection of properties ${\cal L}_1 \cup {\cal L}_2$, together with
all possible products (in the sense that we have explained), as
empirically adequate for the description of the compound physical
system. In other words, the fact that one describes two  physical
systems as a whole does not lead one to consider global experiments
on both objects at once. One can object that a description of a compound
physical system that takes only into account the possible
properties of the subobjects and not of the compound system
as a whole is necessarily incomplete.

The underlying set of the coproduct for our example would be
isomorphic  with the collection
of all ordered pairs of non-zero (closed) subspaces of ${\mathbb C}^2$,
with an
additional global least element pasted at the bottom:
\beq
{\cal L}_1({\mathbb C}^2) \coprod {\cal L}_2({\mathbb C}^2) =
\big\{(M_1,M_2)\ |\ M_i \in {\cal L}_i^0({\mathbb C}^2),\ i=1,
2
\big\}
\uplus \big\{0\big\}
\eeq
For a more profound study of the properties of this structure, we refer
to \cite{Aerts1984,Valck2000,Valck2001}. In general, also in this case
the covering law fails, although all corresponding Sasaki projections seem to
behave regularly, which is possible because orthomodularity is in general not
valid.

The standard quantum physical prescriptions for the
construction of a mathematical representation of a
physical system that is conceived as being made up of
several components, require one to construct the Hilbert
tensor product of the Hilbert spaces corresponding to the
putative subobjects, and the possible selection of an
appropriate closed subspace, to account for the fermionic or bosonic nature of
these constituents. It is clear from our analysis that this procedure is not
possible for the case of two separated spin 1/2 particles. On the 
other hand, the
tensor product procedure can be justified in the axiomatic approach, 
given the fact
that the putative compound system satisfies the standard prescriptions of the
axiomatic approach
\cite{Valck2000,AD1978b,Coecke2000}.

Last but not least, we think that the standard notion of so-called
``identity of elementary physical objects in a compound system", which
is so problematic at a fundamental conceptual level, is not
particularly problematic in our approach. Indeed, there may not be such thing
as a physical system consisting of two identical subobjects. Indeed, such a
system may have to be considered as one global physical system,
that may even manifest itself at spatially separated
regions, a problem that would more properly be related to our {\it
a priori}, possibly macroscopically biased, ideas on localization in 
space. Indeed,
experimental evidence suggests that the property of being localized 
in space, is in
general not a classical property (see \cite{Piron1990} and references 
therein). In
that case, the putative compoundness would be a mental construction that is
ascribed in retrospect to the physical system before it actually 
interacted with a
suitable measuring device.

\end{document}